\documentclass[preprint,showpacs,preprintnumbers,amsmath,amssymb]{revtex4}

\usepackage{graphicx}
\usepackage{dcolumn}
\usepackage{bm}

\newcommand{\A}{$\alpha$}
\newcommand{\Ap}{$\alpha^{\prime}$}

\newcommand{\degree}{$^{\circ}$}
\newcommand{\Ex}{$E_x$}
\newcommand{\thcm}{\theta_{c.m.}}

\newcommand{\thlab}{\theta_{\mathrm{lab}}}

\newcommand{\vecr}{\mathbf{r}}

\newcommand{\vecrp}{\mathbf{r^{\prime}}}

\begin{document}

\preprint{APS/123-QED}

\title{Isoscalar Giant Resonance Strengths in $^{32}$S \\
and possible excitations of superdeformed and $^{28}$Si + $\alpha$ 
cluster bandheads}

\author{
M.~Itoh$^1$\footnote[1]{Present address : Cyclotron and Radioisotope
Center, Tohoku University, Sendai, Miyagi 980-8578, Japan}, 
S.~Kishi$^2$, H.~Sakaguchi$^2$, 
H.~Akimune$^3$, M.~Fujiwara$^1$, U.~Garg$^4$, K.~Hara$^1$, 
H.~Hashimoto$^1$, J.~Hoffman$^4$, T.~Kawabata$^5$, K.~Kawase$^1$,
T.~Murakami$^2$, K.~Nakanishi$^1$, B.~K.~Nayak$^4$, 
S.~Terashima$^2$, M.~Uchida$^1$, 
Y.~Yasuda$^2$, M.~Yosoi$^2$
}
\address{
$^1$ \it Research Center for Nuclear Physics (RCNP), Osaka University, 
Osaka 567-0047, Japan \\
$^2$ \it Department of Physics, Kyoto University, 
Kyoto 606-8502, Japan \\
$^3$ \it Department of Physics, Konan University, Hyogo 658-8501, Japan \\
$^4$ \it Physics Department, University of Notre Dame, 
Notre Dame, IN 46556, USA \\
$^5$ \it Center for Nuclear Study, University of Tokyo, 
Wako, Saitama 351-0198, Japan \\
}

\date{\today}

\begin{abstract}
Isoscalar giant resonances and low spin states 
in $^{32}$S have been measured with inelastic $\alpha$ 
scattering at extremely forward angles including zero degrees at 
E$_{\alpha}$ = 386 MeV. 
By applying the multipole decomposition analysis, various excited states are 
classified according to their spin and parities (J$^{\pi}$), 
and are discussed in relation to the super 
deformed and $^{28}$Si + $\alpha$ cluster bands. 
\end{abstract}

\maketitle

\section{\label{sec:Introduction}Introduction}

Giant resonances of nuclei are a clear manifestation of the strong 
collective excitation modes in many-body quantum systems. 
Detailed experimental and theoretical studies have been devoted to find out 
all possible giant resonances with various multipole 
transitions \cite{harakeh-wounde01}. 
Inelastic $\alpha$ scattering has been used as 
the most suitable tool 
to extract isoscalar multipole strengths. Since $\alpha$ particle 
has  $S=0$ and $T=0$ and the first excited state is as high as 20.2 MeV, 
only isoscalar natural parity transitions are strongly excited, 
(an exception is the weak Coulomb excitation 
of the isovector giant dipole resonance, $T = 1$ and $L = 1$).  
At extremely forward scattering-angles including 0\degree, 
cross sections for states with small transferred angular momenta ($L$) are 
strongly enhanced. In addition, ${\alpha}$ angular distributions at high 
bombarding energies are characterized by
clear diffraction patterns. These characteristic features 
allow us to reliably determine the multipole transition strengths. 
In fact, by means of the multipole decomposition analysis (MDA), 
many isoscalar giant resonances have 
been successfully determined, and their excitation strengths have been 
extracted in recent years 
by the RCNP and the Texas A \& M groups~\cite{Uchida-04,itoh-sm-03,Nayak-58Ni-06,Li-Sn-07,itoh-gmr-sm-01,itoh-sm-02,Patel12,tamu-gqr76,tamu-28si-07,tamu-28si-02,tamu-40ca-01,tamu-feni-06}. 

The giant resonances in $^{24}$Mg, $^{28}$Si and $^{40}$Ca have been 
already studied by both groups. 
Among these light nuclei, of special interest are
the giant resonances in $^{32}$S with  proton and neutron numbers of 16. 
Various theoretical models such as mean-field approaches, the shell model, 
and cluster-structure and molecular-resonance points of view, 
have predicted that there must exist well-developed superdeformed (SD) bands 
at high excitation energies in $^{32}$S. 
This interesting prediction is made on basis of 
the concept that when the $^{32}$S nucleus attains a superdeformed shape, 
with the ratio 2:1 for the long and short axes, nucleon number 16 becomes a 
magic number and the advent of a stable SD band is expected 
at high excitation energies~\cite{yamagami00,rodriguez00,kimura04}. 
Also,  $^{32}$S is a key nucleus 
to understand the relation between the SD structure in heavy nuclei and 
the cluster structure in light nuclei. 
The SD bands in many light nuclei, 
such as $^{36,38}$Ar, $^{40}$Ca, and $^{56}$Ni 
\cite{sven2,Svensson-01,Ideguchi-40Ca-01,Rudolph-56Ni-99} have been 
discovered in the last decade. 
Therefore, many experiments have been 
performed~\cite{morita85, curtis96, lonnroth10} in order to 
search for the SD band in $^{32}$S. 
However, no clear evidence for the SD band in $^{32}$S 
has so far been reported. 

In the present work, we report the results 
on the $^{32}$S($\alpha$,$\alpha^{\prime}$) experiment 
at $E_{\alpha}$ = 386 MeV.
We find candidate states that might constitute the SD 
and $^{28}$Si + $\alpha$ cluster bands in $^{32}$S. 

\section{\label{sec:Experiment}Experiment}

The experiment was performed at the Ring Cyclotron Facility of 
Research Center for Nuclear Physics (RCNP), Osaka University. 
The details of the experimental setup and procedure are described
 in Ref.~\cite{itoh-sm-03}. 
Here, we present the brief outline of the experiment 
and procedures specific to the present measurement. 

Inelastic scattering of 386 MeV $\alpha$ particles from $^{32}$S 
has been measured at forward angles 
($\theta_{lab}$ = 0$^{\circ}$ $\sim$ 10.5$^{\circ}$). 
In order to identify low-J$^{\pi}$ values of complicated overlapping states, 
background-free measurements in inelastic $\alpha$ scattering 
at forward angles including 0\degree\ were greatly helpful. 
We used two self-supporting natural sulfur foils with thicknesses of 14.3 
mg/cm$^{2}$ for 0\degree\, and of 15.6 mg/cm$^2$ for finite angles. 
The sulfur target was prepared in the following procedure~\cite{matsubara09}. 
At first, the natural sulfur powder (the abundance of $^{32}$S is 
95.02\%) was melted at the temperature of 112.8\degree C. 
The liquid sulfur was solidified between a couple of the Teflon sheets 
with a well defined thickness. 
The target was kept cool during the measurement with liquid 
nitrogen by using the target cooling system described in 
Ref.~\cite{kawabata01} to avoid subliming the sulfur. 

Inelastically scattered \A\ particles were momentum analyzed in 
the high resolution spectrometer, GRAND RAIDEN~\cite{GRaiden}, 
and detected in the focal-plane detector system consisting of 
two multi-wire drift-chambers and two plastic scintillators. 
The scattering angle at the target and the momentum of the scattered 
particles were determined by the ray-tracing method. 
The energy spectra have been obtained in the range of 5 $\leq E_x \leq$ 52 MeV 
at $\thlab$ = 2.5\degree\ $\sim$ 9\degree\ 
and of 6 $\leq E_x \leq$ 50 MeV at 0\degree. 
Measurements were performed  with two different energy-bite settings at 
each angle. 
In the 0\degree\ measurements, the primary beam was stopped 
at the just behind of the D2 magnet of GRAND RAIDEN for the high 
excitation energy bite and at the downstream of the focal-plane 
detector system for the low excitation energy bite. 
At forward angles from 2.5\degree\ to 5\degree, 
the beam was stopped at the location just after the Q1 magnet. 
At the backward angles over 
6.5\degree, the beam was stopped 
in the scattering chamber of GRAND RAIDEN. 
The energy resolution was less than 200 keV through all the runs. 

\begin{figure}
\includegraphics[width=160mm]{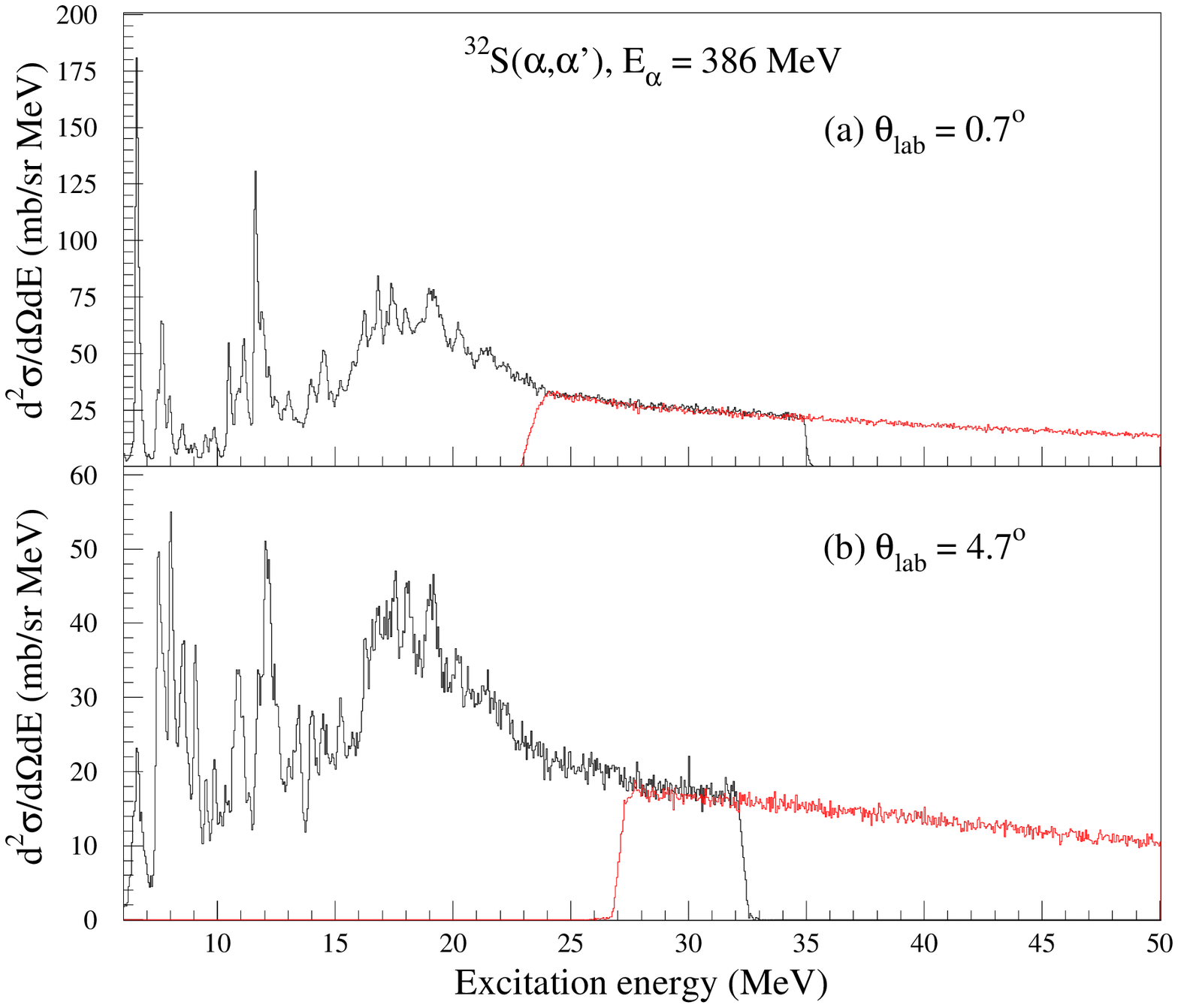}
\caption{(Color online) Excitation energy spectra 
for $^{32}$S(\A,\Ap) at averaged 
laboratory angles of $\thlab$ = 0.7\degree\ and $\thlab$= 4.7\degree. 
The black line shows the energy spectrum obtained 
from the low excitation measurement. 
The red line shows that obtained from the high excitation measurement. 
}
\label{fig:espec}
\end{figure}

Figure~\ref{fig:espec} shows typical 
energy spectra at $\thlab$ = 0.7$^{\circ}$ and 4.2$^{\circ}$. 
In the forward angle measurements, especially at 0\degree, 
backgrounds due to the beam halo and multiple 
Coulomb-scattering become very large. 
However, we eliminated practically all the backgrounds using 
the double-focus property of the ion-optics of the GRAND RAIDEN spectrometer, 
though the effect of the multiple Coulomb-scattering was  smaller 
in the $^{32}$S(\A,\Ap) measurement than those in heavier nuclei such as
$^{208}$Pb. 
Elastic scattering from $^{32}$S was also measured at  
$\thcm$ = 4\degree-27\degree\ to determine the nucleon-\A\ 
interaction parameters with the same incident energy. 

\section{\label{sec:Analysis}Analysis}
\begin{figure}
\includegraphics[width=160mm]{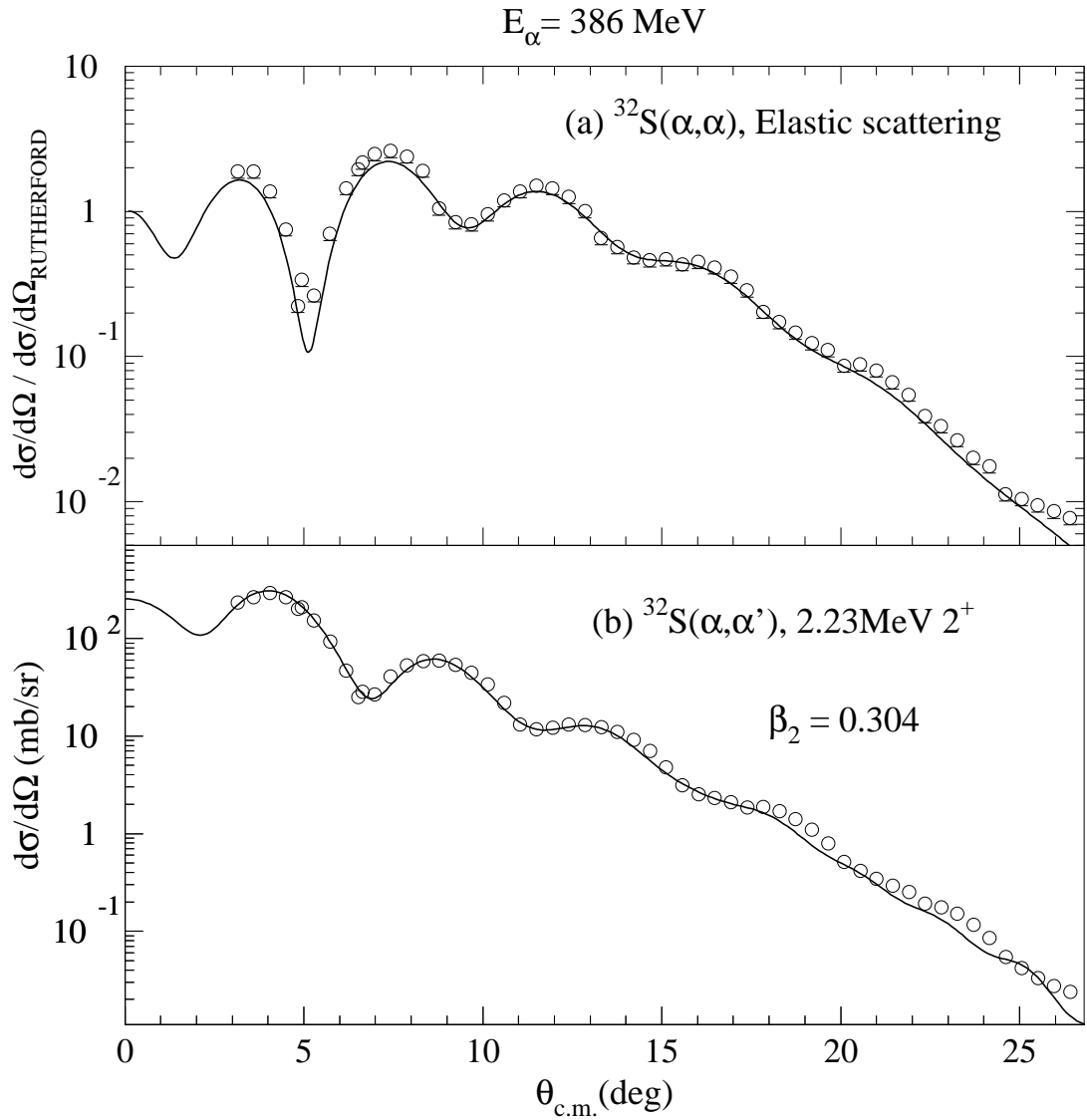}
\caption{(a) Angular distribution of the ratio 
of the differential cross section for elastic scattering 
to Rutherford scattering for 386 MeV \A\ particles from $^{32}$S.  
(b) Angular distribution of 
differential cross sections for the 2.23 MeV 2$^+$ state. 
In both cases, the solid lines show the results of the DWBA calculations 
using the single-folding model (see text). 
}
\label{fig:discrete}
\end{figure}

The MDA has 
been carried out to extract multipole transition strengths from 
E0 to E3, by taking into account the transferred angular momentum ($L$) 
up to $L$ = 13 and minimizing the chi-square per 
degree of freedom. 
$L$ $\geq$ 5 strengths were assumed to be backgrounds due to 
other physical processes such as quasielastic scattering 
in the ($\alpha$,$\alpha^{\prime}$) reaction. 
The cross section data were binned in 
1 MeV energy intervals to reduce the fluctuation effects of the 
beam energy resolution. 
The experimentally obtained angular distributions,
 $\sigma^{exp}(\theta_{c.m.},E_x)$, have been fitted by means of the 
least square method with a linear combination of the calculated 
distributions, $\sigma^{calc}(\theta_{c.m.},E_x)$ defined by 
\begin{eqnarray}
\sigma^{exp}(\theta_{c.m.}, E_x) &=& 
\sum_{L} a_L(E_x)\sigma^{calc}_L (\theta_{c.m.}, E_x),
\label{eqn:mda}
\end{eqnarray}
where $a_L(E_x)$ is the energy weighted sum rule fraction for the $L$ 
component. 
In the DWBA calculation, a single-folded potential model was 
employed, with a nucleon-$\alpha$ interaction of the density-dependent 
Gaussian form, as described in Refs.~\cite{satchler2,kolomiets00}. 
The nucleon-$\alpha$ interaction parameters are given by:
\begin{eqnarray}
V(|\vecr - \vecrp|,\rho_0(r'))& =& 
- V (1+\beta_V\rho_0(r')^{2/3}) \exp(-|\vecr-\vecrp|^2/\alpha_V) \nonumber \\ 
& & - i W(1+\beta_W\rho_0(r')^{2/3})\exp(-|\vecr-\vecrp|^2/\alpha_W), 
\label{eqn:ddpint}
\end{eqnarray}
where the ground state density $\rho_0(r^{\prime})$ was obtained 
using the point nucleon density unfolded from the charge density 
distribution~\cite{chargedens}. 
The parameters $V$, $W$, $\alpha_{V,W}$, $\beta_{V,W}$ 
in Eq.~(\ref{eqn:ddpint}) 
were determined by fitting the differential cross sections of 
elastic \A -scattering measured for $^{32}$S at $E_\alpha$ = 386 MeV; 
the fit is shown in Fig.~\ref{fig:discrete}, 
and the obtained parameters are presented in Table~\ref{tab:interaction}. 
The value $\beta_{V,W}$ = -1.9 was adopted from Ref.~\cite{Satchler83}. 
The angular distribution of the 2.23 MeV 2$_1^+$ state was well
reproduced with the known value of $\beta_2$ = 0.304~\cite{tamu-gqr76}. 
Contribution from the isovector giant dipole (IVGDR) component, arising from 
the Coulomb-excitation, was subtracted above the excitation energy of 10 MeV 
by using the gamma absorption cross section~\cite{s32-g-abs}. 
In the $E_x$ $>$ 40 MeV region, IVGDR
strength was approximated by the tail of the Breit-Wigner function to 
smoothly connect to the $E_x$ $\leq$ 40 MeV region. 

\section{\label{sec:Results}Results}
\begin{figure}
\includegraphics[width=160mm]{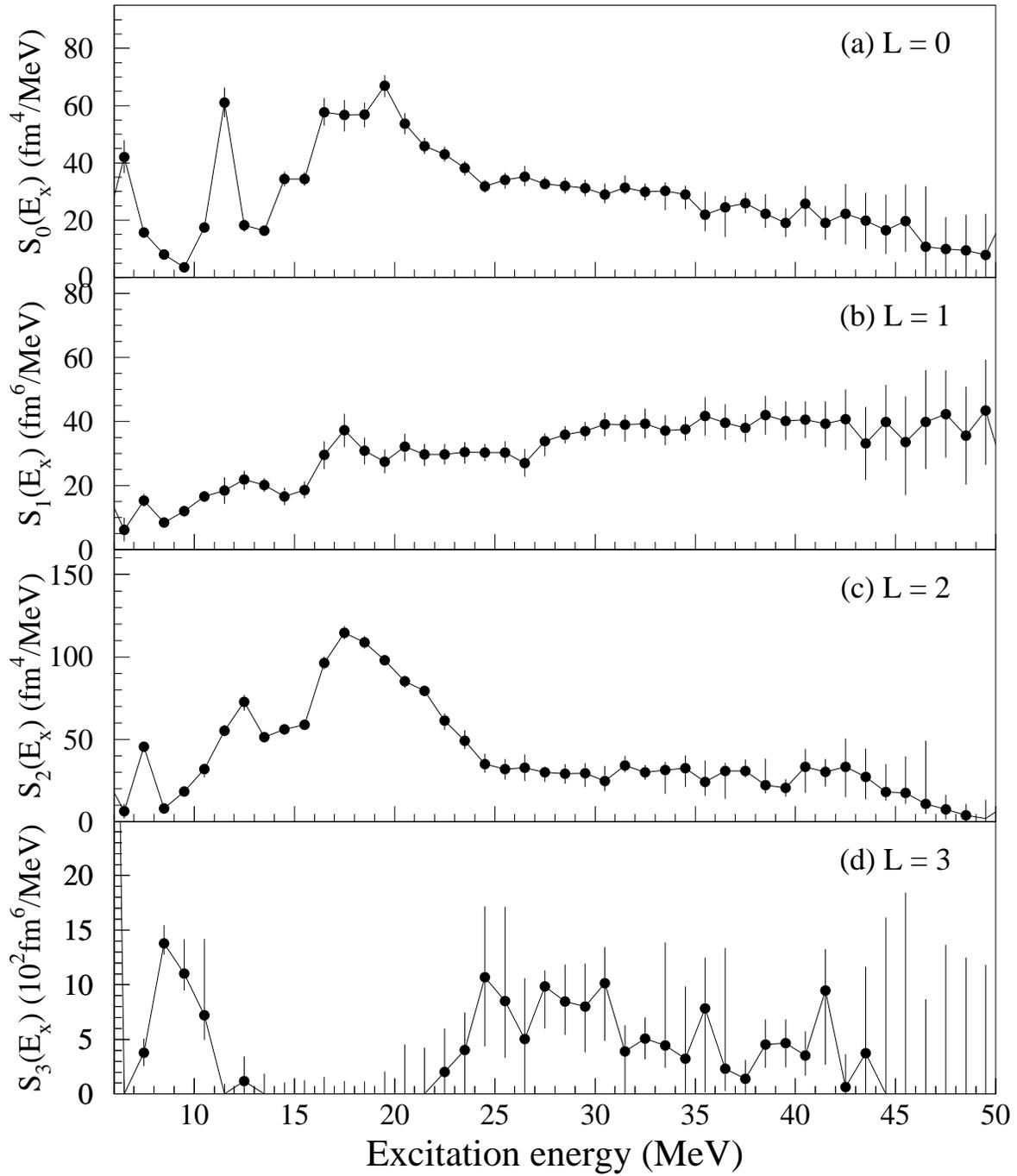}
\caption{Distributions with 1 MeV bins 
for the E0, E1, E2, and E3 strengths are presented in panels (a), (b), (c), 
and (d), respectively.  The lines are given for the guide of eyes. 
}
\label{fig:strength}
\end{figure}

\begin{figure}
\includegraphics[width=160mm]{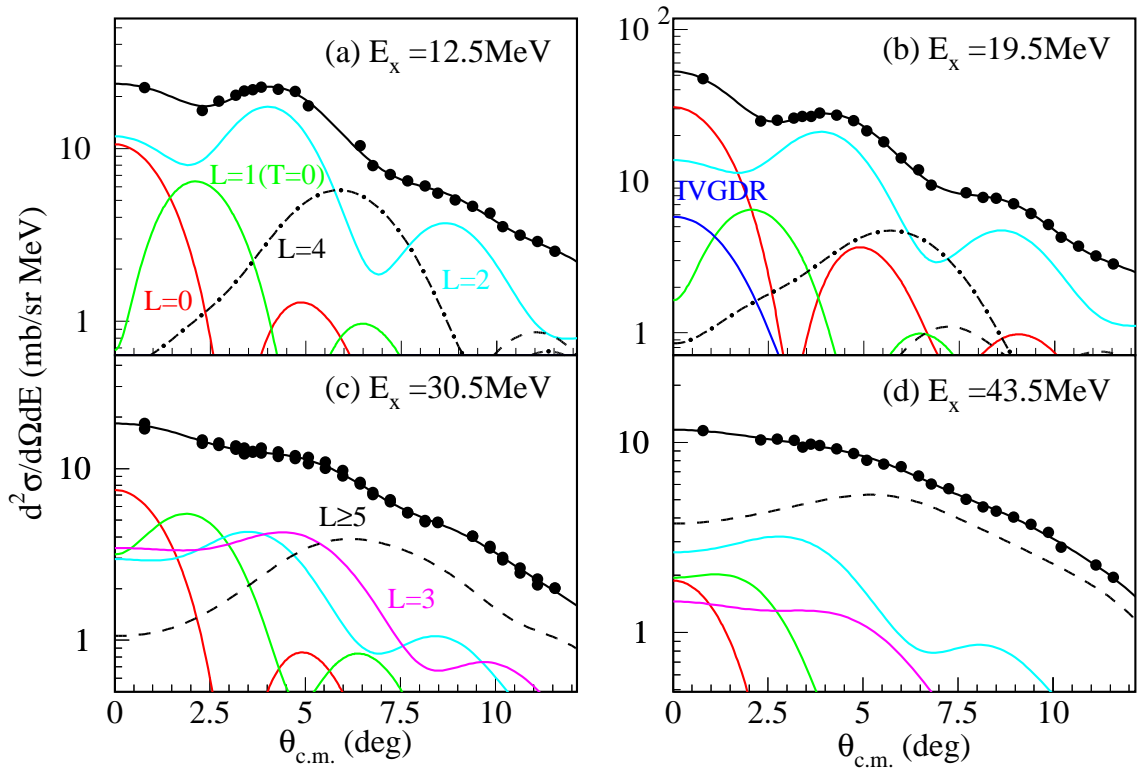}
\caption{(Color online) Typical angular distributions of 
inelastic $\alpha$ scattering.  
The line through the data shows the sum of various multipole components 
obtained by the MDA. 
Each multipole contribution is represented by the color line and 
the transferred angular momentum L are indicated. 
The blue line shows a contribution of the IVGDR estimated from 
the gamma absorption cross section (see text). 
}
\label{fig:grmda}
\end{figure}

Figure~\ref{fig:strength} shows strength distributions for the 
$L$ = 0 (isoscalar giant monopole resonance, E0), 
$L$ = 1 (isoscalar giant dipole resonance, E1), 
$L$ = 2 (isoscalar giant quadrupole resonance, E2), 
and $L$ = 3 (high energy octupole resonance, E3) modes. 
Figure~\ref{fig:grmda} shows typical fitting results of the MDA. 
In the region above $E_x$ = 43.5 MeV, the sum of 
$L$ $\geq$ 5 components constituted dominant part of the cross section
as shown at the right lower part of Fig.~\ref{fig:grmda}. 
Therefore, energy-weighted sum rule (EWSR) values, centroid energies, 
and r.m.s. widths for E0, E1, and E2 have been obtained by summing up from 
6 to 43 MeV. Errors were estimated by changing the summing region to 
$\pm$ 2 MeV (6-41 MeV and 6-45 MeV). 

A total of 108 $^{+7}_{-8}$\% of the E0 EWSR was found. 
The E0 centroid energy (m1/m0) is 23.65 $^{+0.60}_{-0.66}$ MeV, and 
the rms width is 9.43 MeV. 
The isoscalar E1 EWSR fraction is 103 $\pm$ 11\%. However, 
the isoscalar E1 strength continues up to $E_x \sim 50$ MeV, 
similar to that in $^{28}$Si~\cite{tamu-28si-07,madhu}. 
The E2 strength was identified with 143 $^{+9}_{-12}$\% of the EWSR. 
The E2 centroid energy is 22.42 $^{+0.65}_{-0.83}$ MeV, and the rms
width is 9.14 MeV. 

The sum of the E3 strength between 6 MeV and 50 MeV was found to
correspond to  only 33 $^{+7}_{-5}$\% EWSR. 
However, the low excitation energy part between 6
and 18 MeV comprises about 3\% of the EWSR which is equal to that reported 
in $^{28}$Si. 
It would appear that the high energy E3 (HEOR) strength
between 18 and 43 MeV could not be 
separated from higher multipole ($L$ $\geq$ 4) components. 
The centroid energy of the HEOR is 31.4 $^{+0.5}_{-1.0}$ MeV 
which is also comparable to that of $^{28}$Si. 
Although the low excitation energy region of the E4 strength could be 
separated from higher multipole ($L$ $\geq$ 5) components, as described
later, it was not possible to clearly identify the E4 strength 
above \Ex $>$ 25 MeV due to featureless angular distributions, 
as shown in Fig.~\ref{fig:grmda}. 

\begin{figure}
\includegraphics[width=140mm]{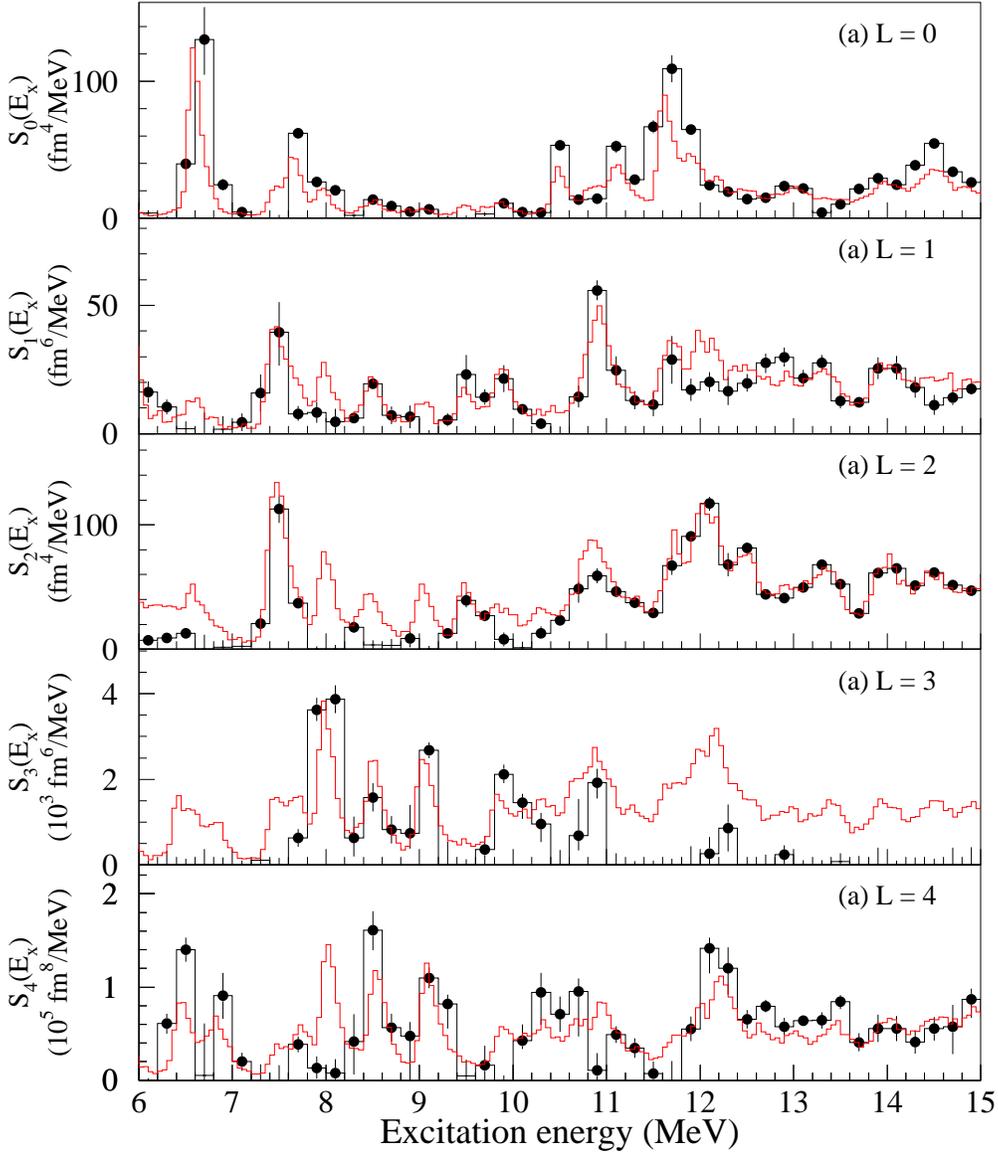}
\caption{(Color online) Strength distributions with a 200 keV bin 
for the E0, E1, E2, E3, and E4 transitions in the region from 6 MeV to 15 MeV 
are presented in the panels (a), (b), (c), (d), and (e), respectively. 
The red lines show the excitation energy spectra 
for the $^{32}$S($\alpha$,$\alpha^{\prime}$) reaction 
at (a) $\theta_{lab}$ = 0.7$^{\circ}$, (b) 2.0$^{\circ}$, 
(c) 3.4$^{\circ}$, (d) 4.8$^{\circ}$, and (e) 5.6$^{\circ}$, respectively, 
scaled to fit in the figures. 
Some differences of peak positions between the excitation energy spectrum 
at $\theta_{lab}$ = 0.7$^{\circ}$ and the E0 strength distribution 
are arising from primarily an artifact of histogramming. 
The differences, if any, are within the uncertainty of 0.05 MeV 
in peak positions. 
}
\label{fig:strength-sd}
\end{figure}

Figure~\ref{fig:strength-sd} shows the distributions 
for the E0, E1, E2, and E3 strengths 
obtained by the MDA with a small bin size of 200 keV. for 
$L$ = 0, 1, 2, 3, and 4. 
In order to obtain excitation energies of the 0$^+$, 1$^-$, 2$^+$, 3$^-$, 
and 4$^+$ levels, we fitted energy spectra with a Gaussian 
at 0.7$^{\circ}$, 1.9$^{\circ}$, 3.3$^{\circ}$, 4.8$^{\circ}$, and 5.6$^{\circ}$, 
respectively. 
The transition strengths were estimated by integrating 
the strength distribution corresponding to the states. 
It should be noted that their absolute values are strongly 
affected by the DWBA calculation used in the MDA. 
The extracted excitation energies and strengths 
are listed in Table~\ref{tab:positive}. 
In the $L$ = 0 strength distribution presented in Fig.~\ref{fig:strength-sd}, 
there were many candidates for the E0 strength at \Ex\ $<$ 14 MeV. 
However, since the isovector E1 cross section 
due to the Coulomb-force shows also a strong peak at 0\degree\, 
similar to the E0 strength, it could not be excluded from the E0 strength 
at \Ex\ $<$ 14 MeV. 
A possible way to look at the IVGDR contribution is to compare 
the ($\alpha$,$\alpha^{\prime}$) strength distributions 
with those obtained from (p,p$^{\prime}$) at similar energies. 
Such data are available from Ref.~\cite{matsubara09}. 
From a comparison of the 0$^{\circ}$ spectra 
between the ($\alpha$, $\alpha^{\prime}$) and (p,p$^{\prime}$) reactions, 
we identified 
six 0$^+$ states in the E0 strength distribution of 
($\alpha$,$\alpha^{\prime}$) as listed in Tables~\ref{tab:positive} 
and~\ref{tab:negative}. 

\section{\label{sec:Discussion}Discussion}

\subsection{Strength distributions of the giant resonances}
In light nuclei, the isoscalar giant monopole (ISGMR) strength is 
fragmented into the wide excitation energy region, 
as reviewed in Ref.~\cite{harakeh-wounde01}. 
In recent works on $^{24}$Mg, $^{28}$Si, $^{40}$Ca, and $^{48}$Ca
~\cite{kawabata-24mg, tamu-24mg-09, tamu-28si-07,tamu-40ca-01, tamu-48ca-11}, 
a large part of the E0 strength was found over \Ex $\sim$ 20 MeV. 
The E0 strength in $^{32}$S was also found to be 
fragmented in the wide excitation energy 
region from 6 MeV to 43 MeV as shown in Fig.~\ref{fig:strength}(a). 
The E0 centroid energy of 23.65 $^{+0.60}_{-0.66}$ MeV is 
comparable to the empirical expression, E$_{ISGMR}$ $\sim$ 78 A$^{-1/3}$, 
of 24.6 MeV. 

As for the centroid energy of the isoscalar giant dipole resonance (ISGDR), 
the E1 strength continues up to \Ex $\sim$ 50 MeV, as described 
in the previous section. 
The empirical expression of E$_{ISGDR}$ $\sim$ 133 A$^{-1/3}$ 
found in Ref.~\cite{Uchida-04} becomes 41.9 MeV. 
Although the E1 strength was found almost 100\% in this measurement, 
and since the absolute value of the strength is strongly affected by the 
DWBA calculation used in the MDA, 
it implies the measurement up to the sufficiently high excitation 
energy region is needed to find the whole strength of the ISGDR 
in light nuclei such as $^{32}$S. 

\subsection{Candidate for the bandhead of the SD band in $^{32}$S}
The 0$^+$ states at \Ex = 10.49 MeV, 11.62 MeV, 11.90 MeV are 
candidates for the bandhead state of the SD band. 
The bandhead 0$^+$ state of the SD band in $^{32}$S is predicted 
to appear at \Ex\ = 10 $\sim$ 12 MeV in the HF and HFB 
frameworks~\cite{yamagami00,rodriguez00}. 
It has also been shown that this SD band is essentially identical to 
the Pauli allowed lowest $N$ = 24 band of the $^{16}$O+$^{16}$O 
molecular structure~\cite{kimura04}. 
It is tempting, 
to conjecture that these 0$^{+}$ states might, indeed, be the 
bandhead of a SD band. Extending this conjecture, we observe 
2$^{+}$ and 4$^{+}$ members of the SD band above these excitation energies. 

\begin{figure}
\includegraphics[width=160mm]{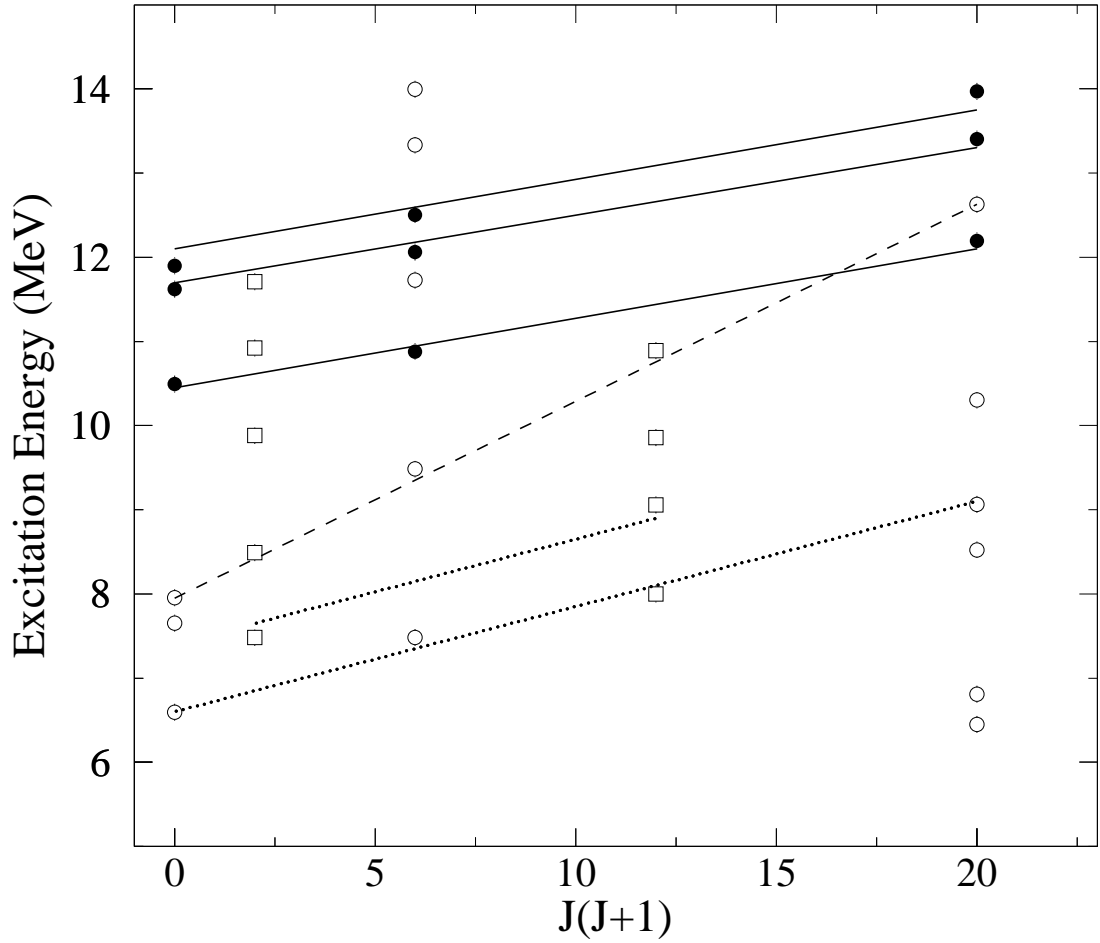}
\caption{
Excitation energies are plotted versus J(J+1) for the states 
obtained in this experiment. 
The open/closed circles show the positive parity states. 
The open squares show the negative parity states. 
Candidates for the SD band are plotted in the closed circles. 
The lines are displayed to guide the eyes. 
The rotational constants $k$ corresponding to the solid, 
dashed, and dotted lines are 83 keV, 234 keV, and 125 keV, respectively.  
}
\label{fig:rotational-band}
\end{figure}

Figure~\ref{fig:rotational-band} 
shows the two-dimensional histogram of the 
excitation energies versus the J(J+1) values. 
The solid lines are drown to guide the eye. 
The slope of these lines corresponds to 
$k \equiv \hbar^{2} / 2 \mathcal{J} =$ 83 keV. 
Although this value is larger than predicted one of 48.5 keV 
in Ref.~\cite{rodriguez00}, it is in good agreement with 
a simple calculation of  $k =$ 85 keV obtained by the assumption of 
point masses for a rigid $^{16}$O + $^{16}$O molecular structure 
with the radius, R = 1.1 A$^{1/3}$ fm.  
It is also comparable to $k$ = 82 keV and 69 keV of the SD bands observed in 
$^{36}$Ar~\cite{sven2,Svensson-01} and $^{40}$Ca~\cite{Ideguchi-40Ca-01}, 
respectively. 
However, the experimental bandheads 
of the SD bands in $^{36}$Ar and $^{40}$Ca are at 
low excitation energies (4.33 MeV and 5.21 MeV, respectively) 
in comparison with \Ex\ = 10 $\sim$ 12 MeV in $^{32}$S. 
This high excitation energy of the bandhead might be a reason why the SD 
band has not been observed in $\gamma$-ray spectroscopic 
studies so far~\cite{sven}. 

In a macroscopic analysis of the $^{16}$O + $^{16}$O rainbow 
scattering, it was concluded that the low-spin 0$^+$, 2$^+$, 4$^+$,
and 6$^+$ states of the N = 24 $^{16}$O + $^{16}$O cluster band 
were fragmented~\cite{ohkubo02} and 
in an elastic $^{28}$Si + $\alpha$ scattering experiment, 
many fragmented 0$^+$ states were observed~\cite{lonnroth10}.
Therefore, the 0$^+$ states at \Ex $\sim$ 11 MeV observed in the present work 
could be the candidates of fragmented 0$^+$ states. 

\subsection{$^{28}$Si + $\alpha$ cluster structure}
The lower excitation energy 0$^+$ states, at 6.6 MeV and 7.9 MeV, 
which are near the $\alpha$-decay threshold 
energy in $^{32}$S, are discussed in relation to the bandhead of the 
$^{28}$Si + $\alpha$ cluster band 
in the analogy with the $^{12}$C + $\alpha$ cluster in $^{16}$O 
and the $^{16}$O + $\alpha$ cluster in $^{20}$Ne~\cite{horiuchi72}. 
Since there are mirror configurations of the $^{12}$C + $\alpha$ and 
$^{16}$O + $\alpha$ clusters, these cluster structures lead 
the parity-doublet rotational bands. 
The appearance of a parity-doublet rotational band in the asymmetric 
intrinsic $\alpha$ cluster configurations is also explained 
by a cluster model with a deep potential~\cite{michel98, ohkubo04}. 

The dashed and dotted lines in Fig.~\ref{fig:rotational-band} are 
drawn to point out members of the parity-doublet 
$^{28}$Si + $\alpha$ cluster band in $^{32}$S. 
The rotational constants $k$ corresponding to the dashed and dotted lines 
are 234 keV and 125 keV, respectively. 
The gap energy between the positive and the negative bands 
for the dashed line is almost zero. It indicates the $^{28}$Si + $\alpha$ 
cluster structure in this band has a rigid body. 
The value of 234 keV is in good agreement with a simple calculation of 245 keV 
obtained with the assumption of point masses 
for a rigid $^{28}$Si + $\alpha$ cluster with a radius R = 1.1 A$^{1/3}$ fm 
for $^{28}$Si, and 1.6 fm for the $\alpha$-particle. 
However, these simple calculations of the rotational constant are just trials 
to explain the experimentally-observed rotational constants. 
More realistic theoretical calculations are highly desired 
for the further detailed comparison with the experimental results. 

\section{\label{sec:Summary}Summary}

We have investigated the isoscalar giant resonance 
strengths in the doubly-closed shell nucleus $^{32}$S, with a view 
to search for the possible superdeformed bandhead predicted 
in theoretical calculations. A novel technique was used to prepare 
an enriched $^{32}$S target, and 
the $^{32}$S($\alpha$,$\alpha^{\prime}$) measurements were made 
at extremely forward angles, including 0$^{\circ}$ 
at E$_{\alpha}$ = 386 MeV.
The extracted E0, E1, E2, and E3 strength distributions from MDA
 are similar to those in nearby light nuclei. 
From the MDA with a 200 keV energy bin, 
three 0$^+$ states at 10.49 MeV, 11.62 MeV, and 11.90 MeV 
are extracted.
These three 0$^+$ states would be candidates for 
the bandhead of the SD band in $^{32}$S. 
In addition, the parity-doublet 
$^{28}$Si + $\alpha$ cluster bands have been identified. 
The rotational constants obtained from the level distance for the 
possible rotation states are in good agreement with simple calculations 
with the assumption of point masses for the $^{16}$O + $^{16}$O and 
$^{28}$Si + $\alpha$ cluster structures. 

\begin{acknowledgments}
We would like to thank H.~Matsubara and A.~Tamii for providing us with 
the $^{32}$S(p,p$^{\prime}$) spectrum at 0\degree.  
We would also like to thank 
K.~Matsuyanagi and E.~Ideguchi, and Odahara for fruitful discussions. 
We wish to thank RCNP staff for providing the high-quality 
$\alpha$ beams required for these measurements.
This work was supported in part by JSPS KAKENHI Grant Number 24740139, 
24540306, and the U.S. National science Foundation 
(Grant Nos. INT03-42942, PHY04-57120, PHY07-58100, and PHY-1068192) 
and by the US-Japan Cooperative Science Program of JSPS. 
\end{acknowledgments}




\newpage 

\vspace{10mm}

\clearpage

\begin{table}[!h]
\begin{center}
\caption{The nucleon-\A\ interaction parameters. employed in the 
folding-model potential used in this work. 
$^*$ taken from Refs.~\cite{satchler2,kolomiets00}
}
\label{tab:interaction}
\begin{tabular}{cccccc}
\hline
 &   V   & $\alpha_V$ &   W   & $\alpha_W$ & $\beta_{V,W}$ \\
 & (MeV) &  (fm$^2$)  & (MeV) &  (fm$^2$)  &   (fm$^2$)    \\
\hline
 $^{32}$S &  31.84 & 3.8 & 17.73 & 3.8 & -1.9$^*$\\
\hline
\end{tabular}
\end{center}
\end{table}

\begin{table}
\begin{center}
\caption{Observed positive parity states. 
Excitation energies are obtained by fitting the energy spectra with 
a Gaussian peak shape at 0$^{\circ}$ for the 0$^+$ states, 
3.3$^{\circ}$ for the 2$^+$ states, and 5.6$^{\circ}$ 
for the 4$^+$ states, respectively. 
Uncertainties in the excitation energies are about 0.05 MeV, which includes 
the fitting errors and the energy calibration errors. 
Each error in strengths is estimated from those of 
the integrated strength distributions. 
}
\label{tab:positive}
\begin{tabular}{ccccccccc}
\hline
Ex (MeV) & J$^{\pi}$ & Strength (fm$^{4}$) & 
Ex (MeV) & J$^{\pi}$ & Strength (fm$^{4}$) & 
Ex (MeV) & J$^{\pi}$ & Strength (10$^5$ fm$^{8}$) \\
\hline
6.59  & 0$^+$ & 39.8$\pm$5   & 
7.48  & 2$^+$ & 34.1$\pm$2.7 & 
6.45  & 4$^+$ & 40.2$\pm$2.2  \\
7.65  & 0$^+$ & 14.6$\pm$1   & 
9.48  & 2$^+$ & 17.3$\pm$2 & 
6.80  & 4$^+$ & 22.1$\pm$5.3 \\
7.95  & 0$^+$ & 7.2$\pm$1    & 
10.88 & 2$^+$ & 30.8$\pm$2.6 & 
8.53  & 4$^+$ & 43.5$\pm$5.2\\
10.49 & 0$^+$ & 10.6$\pm$0.6 & 
11.73 & 2$^+$ & 19.3$\pm$1.8 & 
9.06  & 4$^+$ & 38.3$\pm$5\\
11.62 & 0$^+$ & 29.4$\pm$2.4 & 
12.06 & 2$^+$ & 42.3$\pm$2.1 &
10.3  & 4$^+$ & 27.4$\pm$7\\
11.90 & 0$^+$ & 18.7$\pm$2.4 & 
12.51 & 2$^+$ & 16.3$\pm$0.7 & 
12.19 & 4$^+$ & 52.3$\pm$7\\
      &       &              &
13.33 & 2$^+$ & 24.1$\pm$0.9 & 
12.63 & 4$^+$ & 27.4$\pm$2.3\\
      &       &              &
      &       &              &
13.40 & 4$^+$ & 29.8$\pm$2.4\\
      &       &              &
      &       &              &
13.97 & 4$^+$ & 22.2$\pm$4\\
\hline
\end{tabular}
\end{center}
\end{table}

\begin{table}
\begin{center}
\caption{Observed negative parity states. 
Excitation energies were obtained by fitting the energy spectra 
with a Gaussian peak shape 
at 1.9$^{\circ}$ for the 1$^-$ states and 4.8$^{\circ}$ for the 3$^-$ states, 
respectively. The uncertainties in excitation energies are about 0.05 MeV.  
Each error in strengths is obtained from those of the integrated strength 
distribution. 
}
\label{tab:negative}
\begin{tabular}{ccccccccc}
\hline
Ex (MeV) & J$^{\pi}$ & Strength (fm$^{6}$) & 
Ex (MeV) & J$^{\pi}$ & Strength (10$^3$ fm$^{6}$) \\
\hline
7.48  & 1$^-$ & 11$\pm$5   & 
8.0  & 3$^-$ & 1.5$\pm$0.09  \\
8.49  & 1$^-$ & 5.1$\pm$0.6   & 
9.06  & 3$^-$ & 0.68$\pm$0.09 \\
9.88  & 1$^-$ & 6.2$\pm$1.2  & 
9.86  & 3$^-$ & 0.42$\pm$0.05\\
10.92 & 1$^-$ & 19$\pm$1.4 & 
10.89  & 3$^-$ & 0.52$\pm$0.18\\
11.71 & 1$^-$ & 8$\pm$2.1    & 
   &  & \\
\hline
\end{tabular}
\end{center}
\end{table}
\end{document}